# The Life and Work of Marvin Kenneth Simon

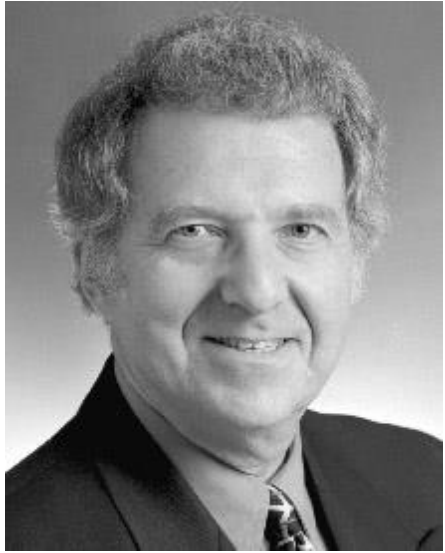

*By M.-S. Alouini, E. Biglieri, D. Divsalar, S. Dolinar, A. Goldsmith, and L. Milstein*

## 1. Introduction

It is a measure of the importance and profundity of Marvin Kenneth Simon's contributions to communication theory that a tribute article and tutorial about his life and work is of current research relevance in spite of the continually accelerating rate of evolution in this area. Marv, as the entire community affectionately knew him, was one of the most prolific and influential communications researchers of his generation. Moreover, he laid the foundation for many of the techniques used in communication systems today. Marv's tragic death on September 23, 2007 continues to engender pangs not only of sadness at the passing of a great friend to many in our community, but also of regret that he is no longer with us to help in resolving the many challenges facing communication systems today.

Marv was born on September 10, 1939, in New York City, NY. He attended Bronx High School of Science, and received the B.S.E.E. degree from the City College of New York in 1960, the M.S.E.E. from Princeton University in 1961, and the Ph.D. degree from New York University in 1966. From l963-l966 he was an instructor in Electrical Engineering at New York University, teaching graduate courses in circuit theory. From l96l-l963 and l966-l968, he was a Member of the Technical Staff at AT&T Bell Laboratories. There he conducted theoretical studies of digital communication systems with a focus on pulse-coded modulation. In 1968 Marv joined the Jet Propulsion Laboratory (JPL), a division of the California Institute of Technology (Caltech) in Pasadena, CA, where he spent the rest of his career, ultimately as a Principal Scientist. In 1978, concurrent with his JPL appointment, Marv began an affiliation with the Electrical Engineering Department at Caltech as a visiting lecturer, which ended in 1996. In this role he created and taught the department's three-quarter sequence of first-year graduate courses on random processes and digital communication. In addition to teaching this sequence a total of ten times over several decades, Marv also mentored many of his Caltech students in their research and careers.

Over four decades starting in 1966, Marv performed seminal research in a wide variety of areas within digital communications and left a deep and broad legacy in each area. For each of these areas, before moving on to the next one, Marv authored or co-authored an influential textbook on the subject, including the fruits of his own research**.** Several of these went on to become the definitive book for students and researchers working in that area. Marv's specific contributions to synchronization, digital modulation, spread spectrum, trellis coding, differential detection, fading channel performance evaluation, and software-defined radios (SDRs) are deep and fundamental; they have had, and will continue to have, a lasting impact on space, terrestrial, satellite and mobile communication system analysis and design. These technical contributions will be described in more detail later in this tutorial and tribute.

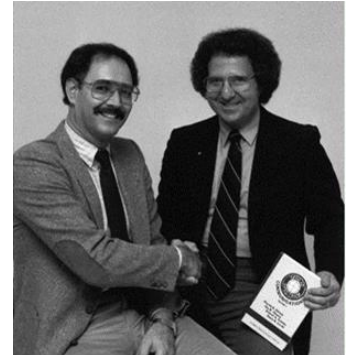

*Marv and Barry Levitt*

Marv's broad and deep contributions to the field of digital communications and leadership in advancing the discipline garnered him much recognition and many awards from both the IEEE and from NASA. He was elevated to IEEE Fellow in 1978 and awarded the IEEE Centennial Medal in 1984 and the IEEE Third Millennium Medal in 2000. In 1997 Marv received the IEEE Communications Society Edwin H. Armstrong Achievement Award, the society's highest honor, recognizing his seminal contributions spanning three decades in the design and analysis of novel coherent digital communication systems. For his outstanding contributions to their deep space and near-earth missions, NASA awarded him its 1979 Exceptional Service Medal and its 1995 Exceptional Engineering Achievement Medal. Marv was also a Fellow of the Institute for the Advancement of Engineering and in 2005 became an IEEE Life Fellow.

Marv published over 200 research papers, several of which were singled out for awards and distinction. In particular, his paper co-authored with Dariush Divsalar, entitled "Multiple symbol differential detection of multiple Phase-Shift-Keying", which appeared in the March 1990 issue of the *IEEE Transactions on Communications*, was selected for inclusion in a book celebrating the fiftieth anniversary of the founding of the IEEE Communications Society. This book, entitled "The Best of the Best: Fifty Years of Communications and Networking Research," contains the 50 research papers that, during the previous five decades, were the most significant and frequently cited in the evolution of communications systems design and analysis over that period. Marv was also the co-recipient of the 1990 *IEEE Transactions on Vehicular Technology* Paper Award for his work on trellis coded differential detection systems, and the co-recipient of the 1999 IEEE Vehicular Technology Conference Paper Award for his work on the performance analysis of dual-branch switched diversity systems over fading channels.

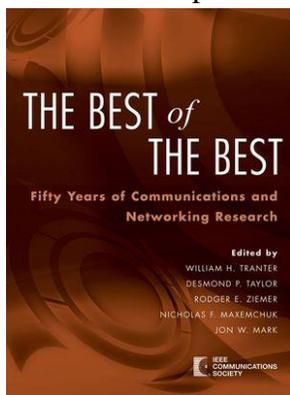

A prolific writer of technical books, Marv authored or co-authored seven: *Telecommunications Systems Engineering* (with Lindsey; Prentice-Hall, 1973, and Dover Press, 1991), *Spread Spectrum Communications*, Volumes I, II, and III (with Omura, Scholtz, and Levitt; Computer Science Press, 1984, and McGraw Hill, 1994, 2$^{nd}$ ed. 2002), *Introduction to Trellis Coded Modulation with Applications* (with Biglieri, Divsalar, and McLane; MacMillan, 1991), *Digital Communication Techniques: Signal Design and Detection* (with Hinedi and Lindsey; Prentice Hall, 1994), *Digital Communication over Fading Channels* (with Alouini; Wiley, 2000, 2nd ed. 2005), *Probability Distributions Involving Gaussian Random Variables: Handbook for Engineers and Scientists* (Kluwer, 2002), and *Bandwidth-Efficient Digital Modulation with Application to Deep Space Communication* (Wiley, 2003). These textbooks were highly cited and widely used by students and practitioners throughout the world. Marv also served as editor of two books on collected works around specific topics: *Phase-Locked Loops and Their Application* (IEEE Press, 1978) and *Autonomous Software-Defined Radio Receivers for Deep Space Applications* (Wiley, 2006); in the latter work he also co-authored most chapters. In addition, he co-authored several chapters in the book *Deep Space Telecommunication Systems Engineering* (Plenum, 1984) as well as chapters on Spread Spectrum Communications in the *Mobile Communications Handbook* (CRC Press, 1995), the *Communications Handbook* (CRC Press, 1997), and the *Electrical Engineering Handbook* (CRC Press, 1997).

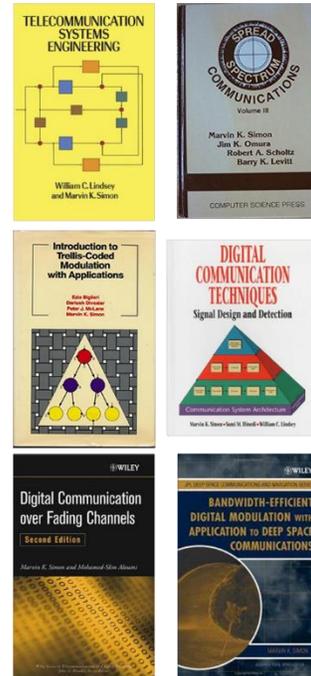

Marv's works were highly cited, often long after publication. Indeed, one reviewer of Marv's 1998 *IEEE Proceedings* paper on performance analysis of digital communications in generalized fading, jointly authored with Slim Alouini, stated, "This paper is a groundbreaking piece of work (that) will have a long and useful life." This was a prophetic statement as Marv's co-authored textbook on this technique is his most cited work, with over 6000 citations in Google scholar. Marv's research as applied to the design of NASA's deep space and near-earth missions resulted in the issuance of 12 U.S. patents, 29 NASA Tech Briefs and 4 NASA Space Act awards. In addition to over 200 published research papers, Marv also contributed to over 80 JPL technical publications. Marv's overall body of work has been cited over 24,000 times in Google Scholar, with an h-index of 57. Since 2010, five years after his passing, Marv's work has been cited almost 10,000 times, a remarkable tribute to the longevity of his contributions and impact.

Marv was very active in the IEEE Communications Society (ComSoc), and a frequent and welcomed presence at its symposia and workshops. Perhaps his most beloved technical event was the IEEE Communication Theory Workshop. Without any published proceedings and with its relaxed format and prestigious participants, this workshop allowed for a free and creative exchange of ideas as well as in-depth analysis and debate. Marv was a constant fixture at the workshop, presenting thought-provoking talks, asking probing questions of other presenters, and taking part in the pervasive informal technical discussions throughout the day and sometimes late into the night.

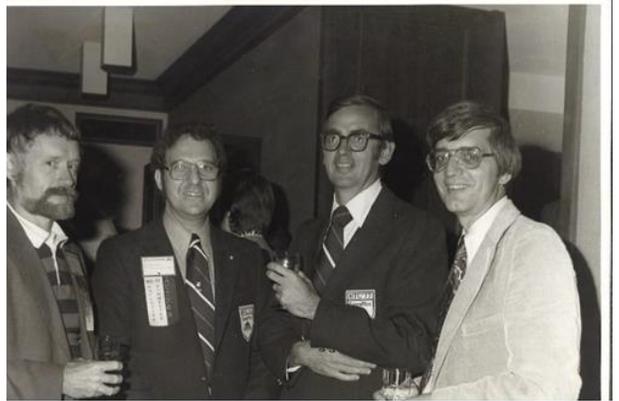

*From Left: Edward Weldon, Marv, Robert Scholtz and Robert Lucky*

In addition to being a leader in the field Marv was, for those who had the good fortune to interact with him, a researcher, professor, collaborator, and mentor. Memories of Marv bespeak a man who took intense interest both in the problems he wished to solve and the people with whom he worked. As a researcher, Marv was intrigued by interesting problems, and smart enough, creative enough, and persistent enough to solve them. As a teacher, Marv provided his students with the fundamental knowledge they needed to form the foundation for their own research. As a collaborator, Marv was an eternal graduate student, always hungry for new discoveries and new results and eager to investigate them deeply. Reflections on these many dimensions of Marv's professional life are provided in the next section.

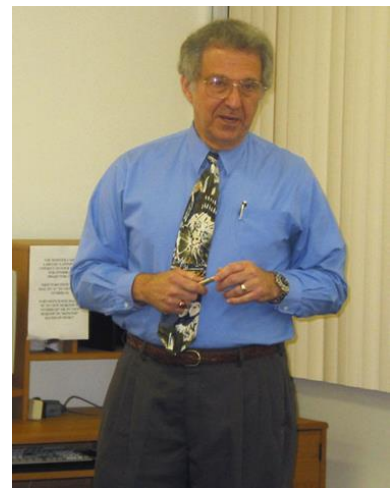

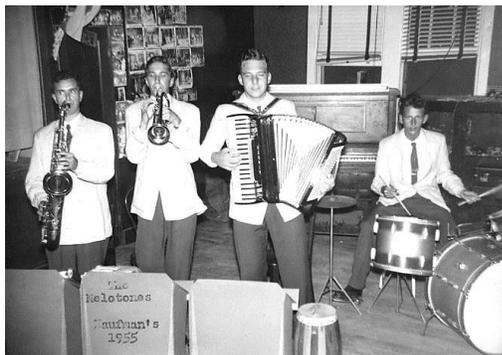

Marv's intellect encompassed topics well beyond research. In his early years Marv was a talented accordion player and performed regularly on a TV variety show called The Horn and Hardart Children's Hour, as well as in the Catskills with his band, "Marvin and the Mellowtones". He eventually chose engineering over music for his career (much to the benefit of our research community), but music remained his avocation. He frequently shared his musical talents with his peers, entertaining his colleagues on the piano with remarkable finesse and talent at workshops, in their homes, or on the concert-sized grand piano that occupied most of his living room. In 2004 Marv accompanied Robert McEliece in his rendition of "Thank you very much" for one of the most memorable Claude E. Shannon Award Lectures ever presented. One of Marv's other beloved hobbies was magic; he would perform tricks for colleagues and at his kids' birthday parties under the moniker "Marvin the Magnificent". Marv was also an award-winning photographer and a prolific author of non-technical books; in the 1980s he became one of the world's experts on computer adventure games, writing several best-selling books on the topic, including "Keys to Solving Computer Adventure Games" and "Hints, Maps, and Solutions to Computer Adventure Games."

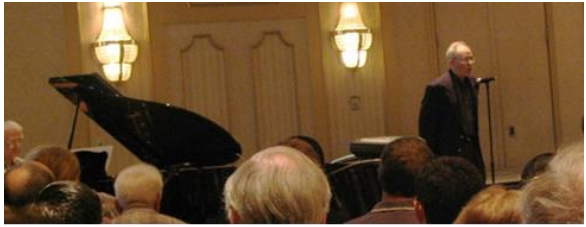

*Marv accompanying Robert McEliece's 2004 Shannon Lecture*

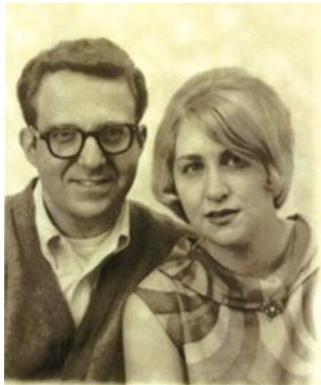

*Marv and Anita*

Marv was also an extraordinary husband and father. He met his future wife Anita at a dance while he was working towards his Ph.D. (which to Anita indicated great husband potential). Marv and Anita were constant companions and deeply in love throughout their 41.5 years of marriage, perfectly counterbalancing each other's character traits. Sadly, Anita passed away just 8 months after Marv's death. Marv was also a doting and loving father, and his pride in his children was palpable. He constantly bragged about his daughter Brette, a highly accomplished corporate lawyer in Southern California, and his son Jeffrey, a nutritional consultant and successful entrepreneur. While Marv's death struck a huge and tragic blow to the research community, the loss to his family was infinitely greater. He is missed by all.

Following the next section reflecting on Marv's many professional dimensions, Marv's dominant technical contributions in the areas of synchronization, spread spectrum, interference cancellation, trellis coding, differential detection, fading channel performance evaluation, and SDRs are described. The significance of these contributions cannot be overstated. In particular, Marv's early pioneering work on carrier and symbol synchronization, tracking and acquisition earned him a reputation as one of the world's leading authorities on this subject.  Moreover, Marv's contributions to time and frequency synchronization of direct sequence and frequency-hopped communications served as the basis for synchronization design and analysis in many military and commercial systems.  Marv together with Divsalar were the first to propose criteria for designing trellis codes to operate over channels with multipath fading.  These criteria have since been employed by many others, most recently for the design of space-time trellis codes. Marv's research on multiple symbol differential detection, also jointly with Divsalar, demonstrated the ability to obtain, with a simple implementation modification, significantly improved performance over the classical (two-symbol observation) method.  The reduction to practice of this theoretical work resulted in the issuance of a landmark patent.

Marv's discovery with Alouini in the late 1990s of the moment-generating function (MGF)-based unified approach to the performance analysis of communication over fading channels stimulated a vast amount of research throughout the communications community. This novel approach to performance analysis drastically simplified previous approaches both analytically and computationally and led to new results that heretofore had resisted solution.  Marv's last significant contributions were in the area of SDR. In that work he developed and tested a variety of classifiers, estimators, and synchronization modules that, in combination, allow a radio receiver to, on its own, adaptively reconfigure itself to its environment based solely on knowledge obtained from the received signal. This technology is significant because it transforms an ordinary SDR, which is simply reprogrammable, or reconfigurable, into something that is completely autonomous and which can react intelligently to whatever kind of telecommunications signal is sent to it. This is extremely important for deep space assets, which are difficult to reprogram or reconfigure because of the distances involved and low uplink data rates. These many profound technical contributions by Marv will be elaborated on in Section 3, followed by closing remarks in Section 4.

## 2. Reflections on Marv Simon

The professional accomplishments of Marv described in the previous section let us know the caliber of the man. The stories in our memories– of Marv as a researcher, collaborator, mentor, professor, and leader in the field – let us know the quality of the person. This section provides collective reflections on these many dimensions of Marv Simon.

### *Marv as a Researcher*

Marv was a consummate researcher, always hungry for new discoveries and results. Interesting ideas would come to him constantly, and he would investigate all of them. His outstanding computational skills made it possible for him to tackle extremely difficult performance analysis problems. Moreover, he strove to obtain closed-form expressions for his results, following Hamming's philosophy that "the purpose of computing is insight, not numbers." To do so he would generate page after page of complex analysis in his own unique stylistic script. He also spent a considerable amount of time "playing" with any newly derived equations that he worked on to put them in the most compact and elegant form. Indeed, he believed that mathematical elegance was important. His objective was not just to end up with an expression that could be computed but rather an expression that was written and presented in such a way that made it easy to understand and allowed the reader to see easily the dependence of a particular performance metric on the key system parameters.

One of Marv's most unique characteristics as a researcher was that he would not only solve the problem at hand, but his creative imagination would spawn a related problem, then another and another, and he'd develop a common theoretical framework that could treat all of them. Pretty soon he'd have a book-length rounded subject matter rather than a collection of random problems and solutions. And he'd always bequeath to his fellow researchers a thorough documentation of his analyses in the form of a book or multiple peer-reviewed papers. He also loved elegance of style, and was hence an early fan of the Mac computer, using it to create some of his later books in camera-ready form.

### *Marv as a Collaborator and Mentor*

The diversity of Marv's collaborations is quite extraordinary, spanning a broad swath of topics and people. In many of his collaborative projects, Marv served as the inspiration and chief motivator. He also worked to ensure that all his joint publications adhered to his high standards of eloquence and style. Marv's speed in writing a paper once the results were obtained was amazing; many times, it took no more than a single day. Though quick, he always took care with his expressions. Marv synthesized ideas quickly, wrote naturally like he spoke, even on the most technical of topics, and demonstrated a beautiful talent with words. Few people could match Marv's speed and elegance in writing, and he served as a benchmark and golden role model for all his collaborators in this regard.

Marv took great pleasure in teaching and mentoring. Whether at Caltech or JPL, he would sit down with novice students and engineers, explaining with great patience and expertise the fundamentals of the problem at hand. Marv worked closely with several of these young researchers, co-authoring important publications with them that helped to launch their careers. In fact, one of his landmark textbooks was written with a researcher just starting his academic career. It is common advice to Assistant Professors that they should concentrate on writing and publishing papers, which will help them establish their research credentials and earn tenure; authoring books, on the other hand, requires significantly more time and effort, often with little payoff. Marv provided exactly the opposite guidance, advising that writing a book would have important reference value and would establish his mentee as a worldwide authority in the field, which was exactly what transpired. Marv cared a great deal about his mentees, and would offer his time and advice for any challenges they might be facing, technical or otherwise. He also promoted his mentees and their work within the professional community to ensure they received recognition for their accomplishments.

### *Marv as a Professor*

In addition to his position at JPL, over ten years Marv served as a visiting lecturer in the department of Electrical Engineering at Caltech. In this role he developed a year-long course on Digital Communication Theory. This was a classic course and a special experience for students in the field at Caltech. Each of the students was challenged by Marv from the first day of class. Although these students were very accomplished with a strong background in communications, Marv showed them what they did not know by going back to basic principles. He taught the course from his own notes rather than a textbook, and relied heavily on probability theory, random processes, and detection and estimation fundamental principles. His lectures were always thought provoking and very lively.

Marv also had an amazing encyclopedic knowledge of the field and, off the top of his head, could tell many anecdotes about a topic, including when it was presented and discussed for the first time, the researchers and groups that first worked on the topic, what groups were collaborating in the area, the evolution of the topic over the years, and practical implementations of the results in NASA/JPL missions. Through his many examples, Marv demonstrated the importance of knowing the work in a field well in order to build on what came before and also see where the field was heading.

All of Marv's assignments and exams were take home, which took away the stress of sitting through an exam but actually required huge amounts of time and creativity. As the students carried the questions home, they knew that they would be challenging but very interesting. In fact, Marv designed many of the problems he posed on the assignments and exams himself, out of recently published papers in the *IEEE Transactions on Communications*. Marv taught not only the material students were supposed to learn but, more importantly, how to be good researchers and how to tackle open problems in the digital communications field. Though he had high expectations and was always demanding, Marv was an excellent teacher who focused on the fundamentals to provide a foundation for his students' research.

### *Marv as a Leader in the Field*

Beyond impacting the lives of individuals, Marv also personally impacted the field, not just through his research but also by his professional service to the community and as a vocal member of research groups and societies. Marv cared a great deal about the well-being of the Communication Theory field. He was a lynchpin of ComSoc's Communication Theory Technical Committee (CTTC), which he joined in 1973 and participated in thereafter, serving as its Chairman from 1977 to 1980, a critical period of resurgent interest in the field. In this role he was responsible for leading the organization and technical contributions of CTTC in all of ComSoc's conferences and workshops. He was also an Editor of the *IEEE Transactions on Communications* in the area of Communication Theory from l973 to l976. Marv served frequently as Chairman and Session Organizer for technical sessions in IEEE conferences including the National Electronics Conference (NEC), the National Telecommunications Conference (NTC), the International Conference on Communications (ICC), the International Symposium on Information Theory (ISIT), and the Global Telecommunications Conference (GLOBECOM). In all of these roles Marv strove to ensure that meetings and publications around the topic of Communication Theory would adhere to the strictest standards of principle and quality.

Marv was a conscientious editor and reviewer, exerting a very strong influence on papers under his purview. When he agreed to review a paper, he typically invested much effort into double-checking all of its mathematical details. He would also devote a significant amount of time to improving the technical content of the paper by suggesting alternative mathematical proofs or asking for specific clarifications. He also often sent the authors of papers he reviewed his own marked-up version of the paper in which he made language-related comments, suggestions for better phrasing and grammatical corrections to improve the presentation and readability of the paper. This was extremely unusual, as most paper reviewers prefer to remain anonymous. While Marv's standard for what was acceptable as a paper was very high, he always helped the authors to improve their work. Moreover, when he was impressed with a particular research paper, he was very generous in his praise of both the paper and the authors.

Marv was known for his high level of integrity and strong sense of fairness. In particular, he would exert significant effort and time to form a detailed and honest appraisal of any work or colleague he was asked to evaluate. Marv was once asked to write a letter of recommendation for the early promotion of a younger colleague. Though he was not completely familiar with this person's work, he took the time to read many of his papers and then scheduled a phone meeting with him to ask him detailed questions about his work before writing his letter. Marv was typically quick with praise when merited, and his criticisms were not only fair, but also often coupled with thoughtful and detailed advice on means of improvement.

# 3. Major Technical Contributions of Marv Simon

## 3.1 Synchronization

During most of the 1970s, Marv's research was dominated by synchronization considerations in digital communications systems. The majority of this work was done jointly with his long-time collaborator, William Lindsey from the University of Southern California (USC). Their work was mostly concerned with carrier synchronization, but also included research on symbol synchronization. In this section, we summarize some of Marv's most important papers in this area.

Starting with the research on carrier synchronization, there were a couple of common themes that pervaded the vast majority of Marv's papers in this area. The first was the concentration on tracking schemes that did not require the use of pilot symbols. There were three synchronization techniques that fell in this category: the multiply-and-divide by $N$ tracking loop, the Costas tracking loop, and the decision-aided loop. The virtue of any of these three loops is that there was no need to waste power (and throughput) by transmitting, in conjunction with the data-bearing waveform, periodically spaced unmodulated pilot tones which can be tracked with a phase-locked loop (PLL). The key downside of the multiple-and-divide loop, as well as the Costas loop, both shown in Figure 1 below, is the need for a nonlinear operation somewhere in the loop to wipe off the data, and this nonlinear operation exacerbates the effect of the thermal noise. Another downside to either of these latter two loops is a phase ambiguity that results from the nonlinear operation, and which has to be removed, often by differential encoding. The downside of the data-aided loop, which operates in a decision-feedback mode, is the standard problem associated with any decision feedback system, namely the fact that a certain percentage of the decisions being fed back will be in error, thus leading to the possibility of a catastrophic failure.

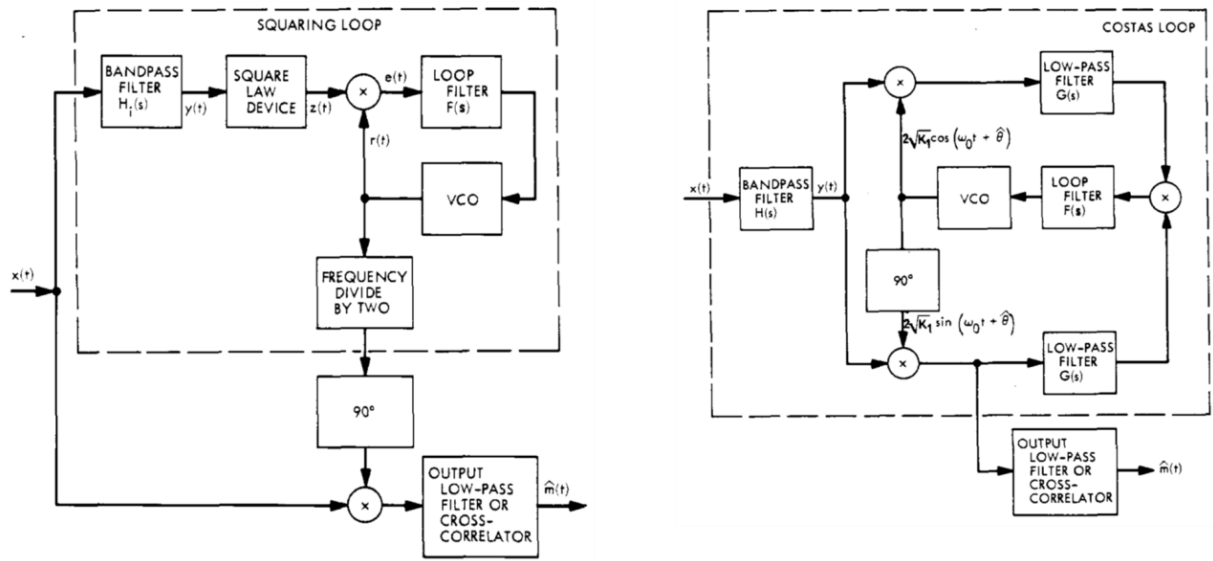

Figure 1 (Figures 1 and 2 of [2]): Second order carrier tracking loops.

The second common thread of many of these papers was the use of a technique by Stratonovich [1] to analyze the system performance. This technique allowed one to approximate the solution to the Fokker-Plank equation (which is the key mathematical equation for studying the dynamics of a PLL) in such a way that a tractable and useful approximation to the probability density function (pdf) of the phase error could be obtained.

One of Marv's early contributions to the theory and practice of carrier tracking loops, which appeared in 1970 in the Proceedings of the IEEE [2], exemplifies both of these common themes. This paper, co-authored with Lindsey, derived the phase error pdf using Stratonovich's technique for the two important tracking loops shown in Fig. 1: the Costas loop and the multiply-and-divide by two loop. Prior work in the literature had focused on analyzing the performance of a first-order loop, whereas in [2] Marv and Lindsey extended that result to these second-order loops. This extension was quite meaningful, as second-order loops were the most commonly used loops at the time, since they are capable of tracking a sinewave with both a phase offset and a frequency offset, and yet they do not have the stability problems that higher order loops have. In addition to deriving the phase error pdf, which can be used broadly in the performance analysis of suppressed-carrier analog and digital modulation, [2] illustrated this use by determining the probability of error for binary phase-shift keying (BPSK) modulation based on the derived pdf.

One of Marv's more heavily cited papers in this area is "Optimum performance of suppressed carrier receivers with Costas loop tracking," which appeared in 1977 and was also co-authored with Lindsey [3]. This work is typical in style of many of Marv's papers: It addresses an important problem, it is mathematically rich, and it contains a plethora of numerical results, a combination that makes the paper very useful to both researchers and practicing engineers. The essence of the paper is to replace an assumption made by many authors that the filters in the in-phase and quadrature branches of a Costas loop do not distort the signal component of the received waveform, meaning they only band-limit the noise. Having removed that idealization, one of the key results of the paper is to show that if carrier synchronization (via the Costas loop) can be performed after symbol synchronization has been achieved, then the passive filters in the two branches can be replaced with active integrate-and-dump filters, and that design change can lead to significant improvements in performance. The performance analysis uses the simplifying assumption that the phase error is small, which allows for linearization. This assumption and hence the performance results hold when the system is in tracking mode but not in acquisition mode.

Other significant papers by Marv and Lindsey on carrier synchronization include [4] and [5]. In the former paper, the tracking of $M$-ary phase-shift keying (MPSK) is studied, with special emphasis on quadrature PSK (QPSK). All three types of carrier-tracking loops, meaning $N$th power multiply-and-divide loops, Costas loops, and data-aided loops, are considered, with numerical results being provided for the probability of error when QPSK is the modulation format. Numerical results are also provided which quantify the loss in effective loop signal-to-noise ratio (SNR) due to the use of an $N$th order nonlinearity to remove the data. In the latter paper, the decision-directed feedback loop is emphasized. A comparison is made between the data-aided loop and a multiply-and-divide by two loop when BPSK is the received waveform, and it is shown numerically that the data-aided loop achieves superior performance.

The final paper on carrier tracking we will survey here was co-authored by Marv and Joel Smith in 1974 [6]. While similar to the other papers mentioned above, in that it makes use of an approximate solution to the Fokker-Plank equation to derive the steady-state pdf of the phase error, and presents numerical results for probability of error, this paper emphasizes a different class of modulation formats, namely offset QPSK and offset quadrature-amplitude shift keying (QASK). The adjective "offset" refers to the fact that the quadrature data streams of the modulated waveform are offset by half a symbol period from one another. This offset has some advantages in systems involving cascades of band-limiting filters followed by a nonlinear amplifier.

Lastly, before leaving this section on Marv's contributions to the field of synchronization, we will summarize a 1977 paper that he and Lindsey wrote in the area of symbol synchronization, as opposed to carrier synchronization [7]. The paper addresses a broad array of topics, including limiting results for both the low and high SNR cases. The authors also derive the maximum a-posteriori (MAP) estimate of the timing epoch, and while the tracking mode of the synchronization loop is emphasized, there is some attention given to the loop's behavior in the acquisition mode. The authors use what is called Manchester coding to ensure that a large run of digits of the same sign will not result in loss of lock, and once again rely on the Fokker-Plank equation as the basis for their analysis.

## 3.2 Spread Spectrum Communications

Marv's best-known contribution to spread spectrum communications was his comprehensive three-volume set of books entitled *Spread Spectrum Communications* [8]. This work, first published in 1985, had Marv as the lead author with Jim Omura, Robert Scholtz, and Barry Levitt as co-authors. The motivation for the book, as stated in its preface, was the widespread interest in spread spectrum emerging at the time, not only for military systems, but also for commercial communications as well as global positioning systems. Indeed, the spread spectrum features of multiple access, rejection of narrowband interference, and averaging of frequency-selective fading helped to make direct sequence spread spectrum the technology of choice in many second and third generation cellular systems, and frequency-hopping the technology underlying Bluetooth systems. *Spread Spectrum Communications* was extremely comprehensive, and spanned topics that, in addition to basic concepts and system models, included algebraic properties of spreading sequences, code division multiple access (CDMA), anti-jam communications, low-probability of intercept systems, and both acquisition and tracking considerations for spread spectrum signals. It dealt with both direct sequence (DS) and frequency hopping (FH), and become one of the most referenced sources on spread spectrum. In 1994 the three volumes were updated, merged, and published as *The Spread Spectrum Communications Handbook* [9], for which a second edition was published in 2002.

Marv's textbook on spread spectrum was the culmination of many years of research on various aspects of this broad topic. His contributions to time and frequency synchronization of direct sequence and frequency-hopped communications make up a large part of Volume 3 in the *Spread Spectrum Communications* book series. He also published many papers on spread spectrum, primarily focused on three topics: acquisition and tracking of the pseudonoise (PN) sequence used in the spreading process, the impact of jamming on system performance, and optimal detection for slow-frequency hopping systems. In spread spectrum communications, the process whereby the receiver synchronizes the phase of its locally-generated spreading sequence to that of the received waveform is done in two steps: The first step is referred to as coarse acquisition, and its goal is to align the two spreading sequences to within plus or minus a large fraction of a chip time. The second process is called tracking, and its goal is to reduce the synchronization error to within plus or minus a very small fraction of a chip time. It is this latter process of PN sequence tracking that Marv's early work on spread spectrum addressed, while his later work focused on code acquisition

There are two classical methods for tracking the phase of the incoming pseudonoise spreading sequence, one that uses a delay-locked loop, and the other, which uses a tau-dither loop. Both of these structures are shown in Figure 2. As shown in the figure, the delay-locked loop has two correlator branches, which operate in parallel, one with a reference signal advanced in phase relative to that of the received signal and the other with the reference signal delayed in phase. The tau-dither loop operates with just a single correlator, but alternates the advanced and delayed reference signals periodically in time via modulation with the binary signal $q(t)$. In essence, both of these loops are variants of an early-late gate. In [10], Marv studies the performance of both loops under the constraint that chip synchronization is to be achieved prior to carrier synchronization, meaning the tracking loops need to operate in a noncoherent manner, since the phase and frequency of the incoming waveform have not yet been removed. Among the contributions Marv makes in [10] is quantifying the performance comparisons between the two loops, and quantifying the effect of using narrowband filters in the correlation branches of the loops.

On the acquisition of spreading codes, Marv teamed up with Andreas Polydoros to develop a generalized serial search code acquisition method [11]. Under this method, they derived analytical expressions for the mean acquisition time, which were then used to optimize the window size of the search strategy. This technique encompassed as special cases the prior work on serial search code acquisition and provided a clean and simple optimization strategy for such algorithms.

A major benefit of spread spectrum for military systems is its ability to mitigate intentional jamming by an adversary. In [12] Marv together with Polydoros studied the impact of partial band noise and tone jamming on two types of quadrature modulation with coherent detection: QPSK and QASK. They derived the worst-case jammer profile and associated worst-case performance for both modulations and show, interestingly, that for high SNR there is a linear dependence between the jammer's optimal power allocation and the system error probability performance. Marv later extended the analysis in [13] to quadrature partial response (QPR) modulation where he showed that the single sample detector for this modulation type is more susceptible to jamming than the matched filter detector of QASK or QPSK. In [14] Marv also studied QASK performance under differentially coherent detection in jamming. This form of detection is more practical than coherent detection under wideband frequency hopping, where the carrier phase is non-continuous from hop to hop. In this work Marv showed that QASK under differentially coherent detection was less susceptible to a jammer than differential phase-shift keying due to the phase sensitivity of the latter modulation to both tone and noise jamming. Marv also developed an analytical method to determine the probability of error for a signal subject to multiple jamming signals in [15]. In this work he develops a recursive solution for the pdf of the squared envelope for a sum of N independent vectors with random phase, where N represents the number of jammers. This pdf can then be used to determine error probability. A similar analysis can be used to obtain the probability of error for a spread spectrum multiple access system, where N represents the number of users.

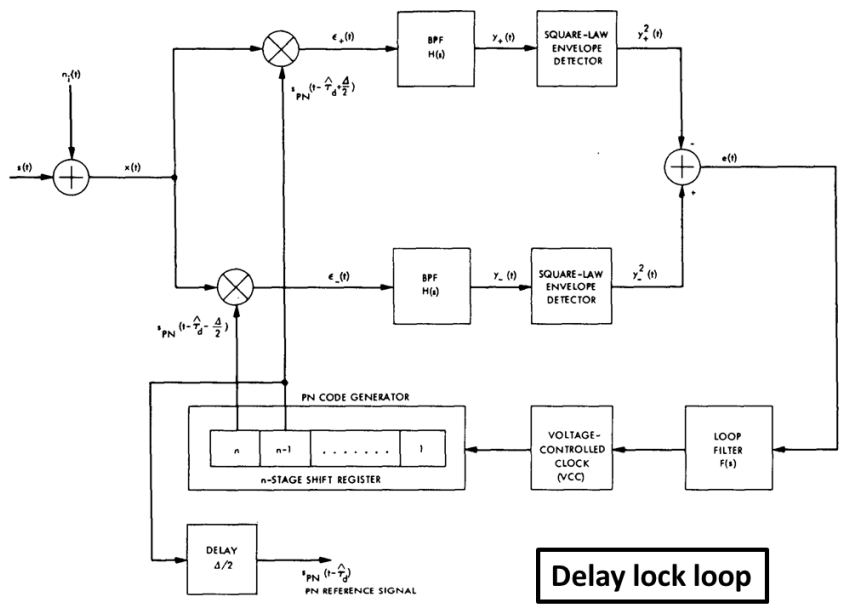

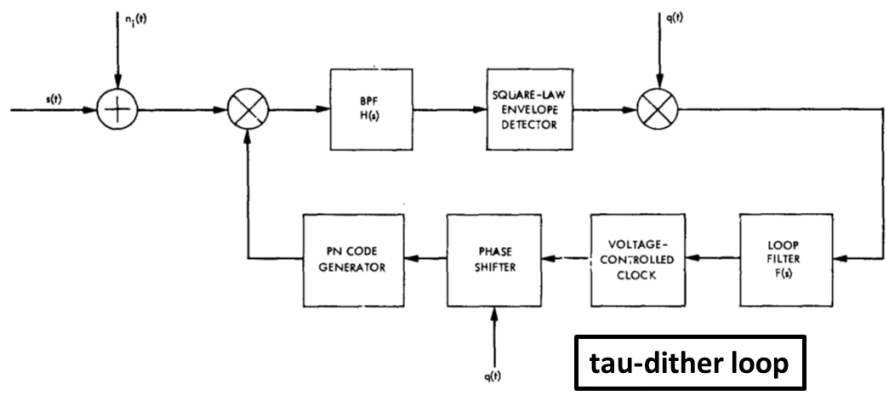

Figure 2 (Figures 1 and 2 of [10]): Delay-locked loop and tau dither loop structures for PN phase acquisition.

Marv's third significant body of work related to spread spectrum, in collaboration with Levitt, Polydoros, and Unjeng Cheng, focused on detection of slow frequency hopped systems [16,17,18]. In such systems there are multiple $M$-ary frequency-shift keying(MFSK) symbols per hop, which makes optimal detection more analytically complex than under fast frequency hopping with a single FSK symbol per tone. The optimal detector for such systems was derived in [16] for both coherent and noncoherent detection based on the average likelihood ratio (ALR) and maximum likelihood ratio (MLR) hypothesis testing methods. This work was extended in [17,18] to develop exact and approximate statistical models based on the ALR and MLR detection methods and using these models to determine the performance of coherent and noncoherent slow frequency hopping MFSK systems.

## 3.3 Multiuser Interference Cancellation

While spread spectrum was first used primarily in military systems for its anti-jam and low probability of detection capabilities, in the late 1980s its multiuser capabilities became of growing interest for commercial cellular systems. Marv's work on interference suppression in spread spectrum multiple access systems was motivated by the fact that in asynchronous CDMA, meaning one where there is no timing synchronization at a given receiver between the arrivals of the various waveforms, the system performance is typically limited by multiple access interference (i.e., interference from the waveforms of the other users in the system). There are multiple ways of combating such interference, and the decision as to what technique to use invariably comes down to a tradeoff between performance and complexity. In particular, most papers in this area endeavor to restrict the complexity to be linear with the number of users. Of these linear complexity techniques, one can attempt to suppress the interfering waveforms either serially or in parallel, and it is the parallel cancellation technique that formed the basis for Marv's primary contribution to multiuser interference suppression.

Beginning in the mid-1990s, Marv together with Divsalar developed a novel parallel interference cancellation scheme that used multiple stages to progressively cancel out more and more interference as the fidelity of the interference estimates improved with each stage [19]. Dan Raphaeli later joined the effort to extend these initial results, as reported in [20]. The inclusion of multistage parallel interference cancellation, in which the amount of interference cancelled is related to the fidelity of the tentative decisions involved in forming the interference estimate, is in general superior to a brute force philosophy of entirely cancelling the interference at each stage. Moreover, the cancellation technique utilized clever nonlinear functions such as a tangent hyperbolic for making the tentative decisions at each stage of the cancellation process, rather than a hard limiter or linear device. The linear device, investigated in [20], has the advantage that the receiver does not require knowledge of the user powers nor does it need carrier synchronization at the various stages. Hence the final data decisions can be performed with a differential (rather than a coherent) detector. The novelty of the general multistage interference cancellation design of [19-20] relative to prior art was recognized with U.S. Patent No. 5,644,592, granted on July 1, 1997.

The parallel partial interference cancellation scheme developed by Marv, Divsalar, and Raphaeli exhibits significantly improved performance over previous parallel and serial processing techniques while retaining linear complexity in the number of users, as compared to the exponential complexity associated with the maximum-likelihood technique of Verdú [21]. In fact, implementation of the scheme requires little additional complexity relative to that already needed to implement a conventional CDMA receiver. Moreover, due to its progressive interference cancellation, the scheme works well for both uniform and non-uniform power distributions of the interference. Interestingly, for single-stage parallel interference cancellation, a uniform user power distribution leads to the best overall performance since each user sees the same amount of interference. In contrast, for successive interference cancellation, a highly non-uniform distribution of user powers, in particular geometric, is desirable so that the strongest user can be reliably subtracted out at each stage [22]. Due to the weighted cancellation based on interference estimate fidelity, at each stage the parallel partial interference cancellation scheme performs similar to successive interference cancellation for highly non-uniform power and similar to parallel interference cancellation for uniform power, getting "the best of both worlds" in terms of performance at each stage for the given residual user power distribution.

The significant performance benefit of the parallel partial interference cancellation scheme over prior techniques is illustrated in Figure 3 from [19]. This plot shows the SNR degradation for different interference cancellation schemes as well as with no interference cancellation, where the degradation is defined as the ratio (in dB) of the SNR required to achieve a given bit-error rate in the presence of other users under the given cancellation scheme versus the SNR required to achieve the same level of performance if only a single user was communicating. The plot assumes a CDMA system operating over an additive-white Gaussian noise (AWGN) channel with equal power users at a bit error rate (BER) of 0.01 and a chip-to-bit rate (spreading gain) of 100. We see from the figure that to support 35 users, conventional CDMA without interference cancellation requires 10 dB additional power per user compared to the case of a single-user system. Using two-stages of partial interference cancellation, the SNR degradation is reduced from 10 dB to 0.7 dB (per simulation results) and, with a third stage, it reduces to 0.25 dB, i.e. the multiuser interference is almost completely removed. The single-stage traditional (brute force) parallel cancellation scheme has an SNR degradation of 1.7 dB, about 1.5 dB higher than the 3-stage partial cancellation scheme. Moreover, the benefit over traditional parallel interference cancellation increases with the number of users: For 80 users the SNR degradation of the 3-stage partial interference cancellation scheme is 1 dB whereas for the traditional parallel interference cancellation scheme it is 8 dB. As shown in [19-20], further minimization of the power requirements can be obtained via coding, which can be seamlessly integrated with the proposed interference cancellation. Hence, Marv and Divsalar demonstrated that the combination of low-complexity partial interference cancellation coupled with channel coding can dramatically reduce the power requirements of each user in a CDMA system, which was the technology used in both third-generation cellular systems as well as 802.11b WiFi.

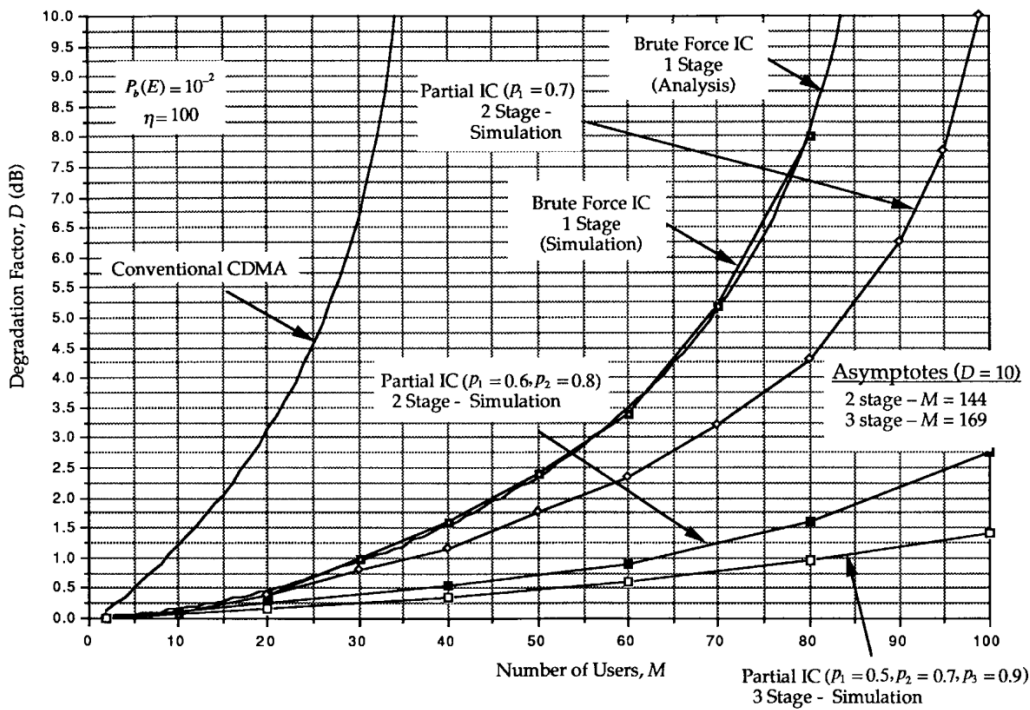

Figure 3 (Figure 6 of [19]): A comparison of the SNR degradation for 2 and 3 stage partial

interference cancellation compared to traditional single-stage full (brute force) interference cancellation. Assumes equal power users, a BER of .01, and a spreading gain of 100.

## 3.4 Trellis-Coded Modulation

In the mid-1980s Marv's attention turned to the area of trellis coded modulation. This field was born in 1974 through a seminal paper [23] written by Jim Massey, which suggested that the performance of a coded transmission system would be improved if modulation and coding were viewed as a combined entity, rather than being designed separately and independently. The most relevant contribution towards the implementation of a system based on Massey's recommendation was the invention of Trellis-Coded Modulation (TCM), described by Ungerboeck in his now-classic 1982 paper [24]. The basic concept was that by augmenting the size of a given signal constellation (typically, by a factor of 2), and using a suitable encoder, one could achieve a larger Euclidean distance than the original constellation (and hence an asymptotic coding gain) and the same transmission rate without increasing the transmitted power or the signal bandwidth. In a power-limited environment, system performance can be improved by using error-correcting codes, which require the modulator to operate at a higher data rate and consequently require a larger bandwidth. In a bandwidth-limited environment, one could use a higher-order signal constellation to increase the bit rate (for example 8-PSK, which carries 3 bits per signal, instead of QPSK, which carries only 2), but this would require a larger power to maintain the same error probability. Now, the TCM solution combines the choice of a higher-order modulation scheme with that of a convolutional code, and, when suitably designed, provides power gain without bandwidth expansion. Thus, any channel that is both power- and bandwidth-limited is ideally suited for TCM.

The original 1982 Ungerboeck paper focused on one- and two-dimensional signal constellations, and was devoted to transmission over the AWGN channel. Fairly soon, extensions to the original designs were sought in several directions, through the use of more general signal classes, more complex designs, and transmission over channels affected by additional impairments. Marv's activity in the '80s evolved along these research lines, described below in some detail and summarized in the textbook [25], published in 1991. Most of this activity was carried out in collaboration with Divsalar, which was described in [25, p. ix] as a "symbiotic working relationship."

### Asymmetric PSK

The original TCM designs (see for example [24]) were based on signal sets with uniformly spaced signal points, as it seemed natural that modulations that are optimum in an uncoded AWGN environment were also optimum for coded systems. This assumption is incorrect, as was shown by Divsalar and Yuen in 1980 [26] for the case of convolutionally encoded QPSK systems. Marv subsequently joined this collaboration to extend the results of [26] to TCM with higher-level PSK, AM, and QAM. Specifically, in [27, 28] Marv together with Divsalar and then with Divsalar and Yuen obtained asymmetric 2M-PSK TCM by augmenting M-PSK with a rotated version of itself. When the rotation $\varphi$ is by $\pi/M$, (symmetric) 2M-PSK is obtained, while rotation values of $\varphi$ other than $\pi/M$ results in an asymmetric *2M*-ary constellation with equal-energy signals. These works also showed that the value of $\varphi$ can be optimized to obtain the largest gain due to asymmetry.

Figure 4 illustrates a simple example of an asymmetric QPSK constellation and the trellis diagram describing a TCM scheme based on a constraint-length 2, rate-1/2 convolutional code. Here the ratio of energies between the in-phase and quadrature components are related to the angle $\varphi$ that describes the asymmetry by $\alpha = \tan^2 \phi/2$. By choosing the "optimum" value of $\alpha$, i.e., one can obtain a tight upper bound to average bit-error probability $P_b$. Comparing this result with upper bounds on probability of error for trellis coded symmetric QPSK and uncoded BPSK yields that the asymptotic power gain is achieved with the TCM design of Figure 4 is 1.76 dB for symmetric QPSK, and 3 dB for the asymmetric signaling.

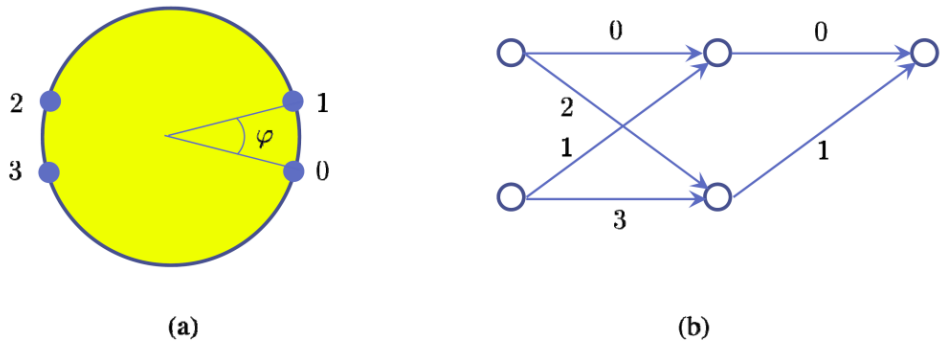

Figure 4: (a) Asymmetric QPSK constellation. (b) Trellis diagram and signal assignment.

## Multiple TCM

A problem arising in certain designs with asymmetric constellations stems from the fact that the largest asymmetry gain turns out to be only achieved in the limit as the points of the constellation merge together. In this case, the TCM scheme becomes catastrophic, i.e., a finite number of channel symbol errors generate an infinite number of decoded bit errors. Even when the signal points do not merge, but are very close to each other, the system becomes much less tolerant of carrier phase synchronization errors. This fact generates the question: How can one achieve some performance gain with respect to "standard" TCM without resorting to modulation asymmetry? One solution makes use of multidimensional constellations, while another technique, advocated by Marv and Divsalar in [29], uses Multiple TCM (MTCM). In its most general form, MTCM is implemented by an encoder whose binary output symbols are mapped into $k$ M-ary symbols in each transmission interval. The parameter $k$ is the multiplicity of the code, and represents the number of M-ary symbols allocated to each branch in the trellis diagram (the choice $k = 1$ yields the original Ungerboeck's designs described in [24]). These designs offer performance gains relative to conventional TCM ($k = 1$) of up to 2 dB with small PSK constellations.

Figure 5 illustrates a simple design. Here there are 4 trellis branches emanating from each state, with two output symbols assigned to each branch. The error event of length 2 in this trellis produces a squared free Euclidean distance equal to 8, independent of the asymmetry angle $\varphi$. The equivalent conventional TCM design based on (symmetric) QPSK achieves a squared free distance equal to 6. More general MTCM designs can achieve performance gains relative to conventional TCM ($k = 1$) of up to 2 dB [25, pp. 267-268].

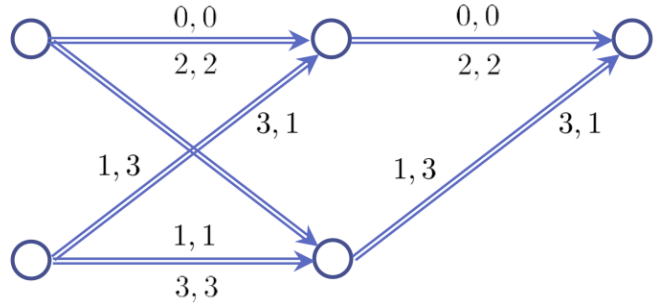

Figure 5: Trellis diagram for rate-2/4 multiple trellis-coded with $k = 2$ and the asymmetric quaternary constellation of Figure 4(a).

### TCM on Fading Channels

Although TCM was originally thought, and designed, for the AWGN channel, so that Euclidean distance was the appropriate parameter to measure its performance, some of its potential application would involve a radio channel, as for example in a mobile satellite system. In addition to its additive noise background, a realistic mobile satellite channel model should include Doppler frequency shift due to vehicle motion, nonlinearities due to a power amplifier, multipath fading, and shadowing. Thus, new design criteria, taking into account the impairments other than additive noise, were called for. Marv and Divsalar, in a sequence of important papers [30, 31, 32, 33], proposed new design criteria for TCM schemes affected by fading and determined their performance. Rician and Rayleigh fading channels were considered, along with coherent and differentially coherent detection, the presence or absence of channel-state information at the receiver, and slow- and fast-fading models. The most important aspect of this body of work was to show that (symbol) Hamming distance rather than the commonly used Euclidean distance is the appropriate design parameter for TCM codes operating over fading channels, particularly in independent (fast) fading. This criterion is used to this day in the design of modern space-time codes.

## 3.5 Multiple Symbol MPSK Differential Detection

Marv's research on multiple symbol MPSK differential detection [34-35], done jointly with Divsalar, demonstrated the ability to obtain, with a simple implementation modification, significantly improved performance over the classical (two-symbol observation) method. The technique makes use of maximum likelihood sequence estimation of the transmitted phases rather than symbol-by-symbol detection as in conventional differential detection over two symbols. As such the performance of this multiple-symbol detection scheme fills the gap between conventional (two-symbol observation) differentially coherent detection of MPSK and ideal coherent detection of MPSK with differential encoding. Figure 6 from [35] shows a parallel implementation of a three-symbol differential detector, where $r_k$ are complex received observations and $\beta_i$ are all possible phases of MPSK. The novelty of these ideas relative to prior art was recognized by the issuance of U.S. Patent 5,017,883 on May 21, 1991.

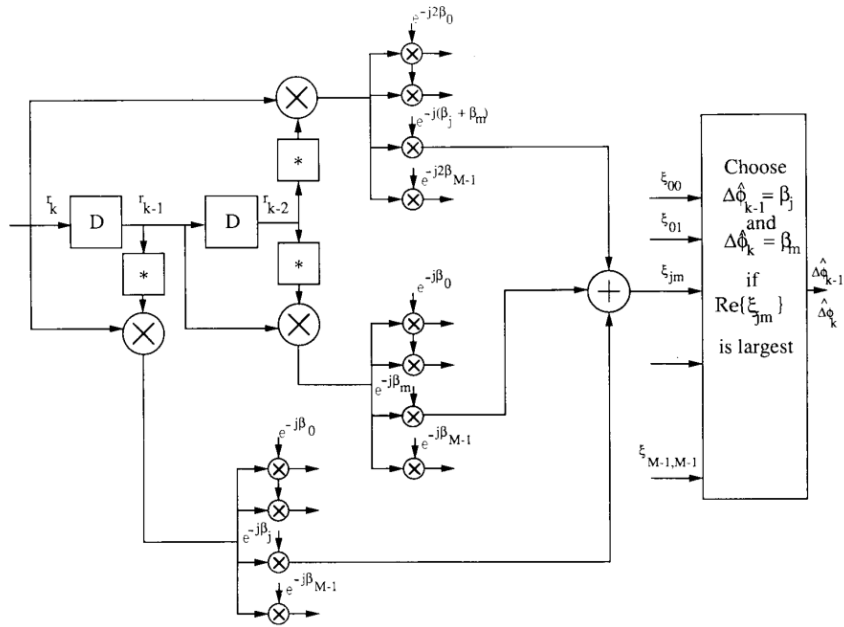

Figure 6 (Figure 5 of [35]): Parallel implementation of three symbol differential detector.

The amount of improvement gained through this technique over conventional differential detection depends on the number of phases *M* in the MPSK modulation as well as the number of additional symbol intervals added to the observation used for detection. What is particularly interesting is that substantial performance improvement can be obtained for only one or two additional symbol intervals of observation. Figure 7 (from [35]) shows the performance results in AWGN using analysis and simulations for *M*=8, and various symbol observation intervals *N*. We see from the figure that at a BER of $10^{-6}$, an SNR of 17 dB is required for the conventional (*N*=2) detector, whereas only 15.8 dB is required for N=3, and that reduces further to a requirement of only 15.1 dB of SNR for *N*=5. The concept was extended to maximum-likelihood differential detection of uncoded and trellis coded amplitude-phase modulation over AWGN and fading channels in [36].

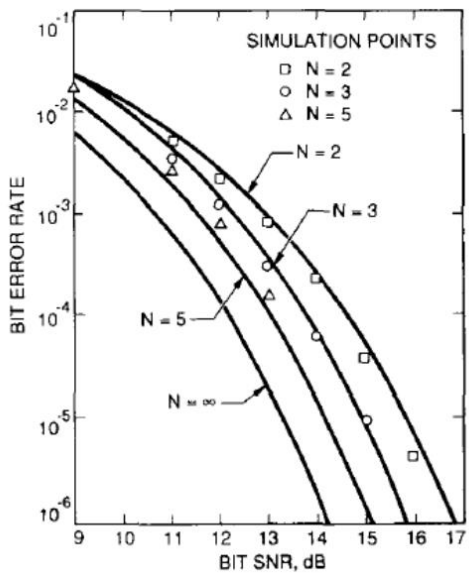

Figure 7 (Figure 6 from [35]): BER performance of multiple symbol differential detection.

## 3.6 Performance of Digital Communications over Fading Channels

Marv's work in the area of digital communication over fading channels can be split into three categories: Derivation of alternative mathematical representations of well-known special functions that facilitate the performance analysis over fading channels, followed by performance studies of uncoded diversity systems over generalized fading channels, which was then extended to include coding. The work on this topic started with the "fascination" of Marv by the alternative representation of the Gaussian Q-function that came out as a byproduct of the study conducted by Craig on the symbol error probability of two-dimensional signal constellations [37]. Together with Divsalar, Marv was able in [38] to capitalize on this new alternative form of the Gaussian Q-function to either simplify or more accurately compute the probability of error of a variety of communication system. More specifically, the error probability of various digital modulations in the presence of Costas loop tracking, fading, carrier synchronization error, and intersymbol interference was treated. This paper offered also a generalization of the one-dimensional alternative representation of the Gaussian Q-function form to the case of a two-dimensional Gaussian Q-function with arbitrary correlation which was used to evaluate exactly the symbol-error probability of MPSK modulation with In-phase/Quadrature-phase unbalance as well as quadrature amplitude modulation.

Motivated by the computational advantages provided by these two new representations of the one and two-dimensional Gaussian Q-functions, Marv discovered a new form of the first-order and generalized Marcum Q-function [39]. His new alternative form extended the simplifications that were made possible with the Gaussian Q-function for the coherent communication systems to the differentially coherent and noncoherent ones. It also enabled the derivation of simple and strict upper and lower bounds of the Marcum Q-function which turned out to be quite useful for the simple evaluation of a tight upper bound on the average bit-error probability performance of a wide class of noncoherent and differentially coherent communication systems operating over generalized fading channels [40]. The new representation of the Marcum Q-function was also quite handy in deriving a new and computationally efficient expression for the bivariate Rayleigh cumulative distribution function (CDF) in the form of a single integral with finite limits and an integrand composed of well-known and simple functions [41].

Marv then began collaboration with Alouini to further study the advantages provided by these various alternative representations. That collaboration resulted in a landmark paper published in the Proceedings of the IEEE presenting a unified approach to evaluating the error-rate performance of digital communication systems operating over a generalized fading channel [42]. This paper analyzed performance of general diversity systems, as shown in Figure 8, for a wide range of coherent, differentially coherent, and noncoherent modulations with different fading distributions associated with each diversity branch. That work showed that in many cases, average error-rate expressions for communication schemes operating over generalized fading environments can be put in the form of a single integral with finite limits and an integrand composed of elementary functions, thus readily enabling numerical evaluation. Subsequently, the performance of many classical diversity-combining schemes was addressed or revisited [43-53]. For instance, a simple, generic, and useful analytical performance analysis and optimization of switch-and-stay diversity systems which have the advantage of offering one of the least complex solutions to mitigating the effect of fading was offered in [43]. The impact of fading type and severity, unbalanced branches, fading correlation, and imperfect channel estimation on the performance of these systems were carefully studied. In addition, a comparison with the performance of more complex diversity schemes such as maximal-ratio combining and selection combining was presented. Looking at another low-complexity diversity scheme, [44] borrowed the notion of the "spacing" between ordered exponential random variables from the Order Statistics literature to develop an elegant performance analysis of the generalized selection combining (GSC) diversity scheme over Rayleigh fading channels. More specifically, starting with the moment generating function (MGF) of the GSC output SNR, closed-form expressions for the average combined SNR, outage probability, and average error probability of a wide variety of modulation schemes operating over independently, identically distributed (i.i.d.) diversity paths were derived. While [44] presented also results for the non-i.i.d. case, these initial results were still quite complicated and were therefore simplified and put in a more compact and elegant form in [45].

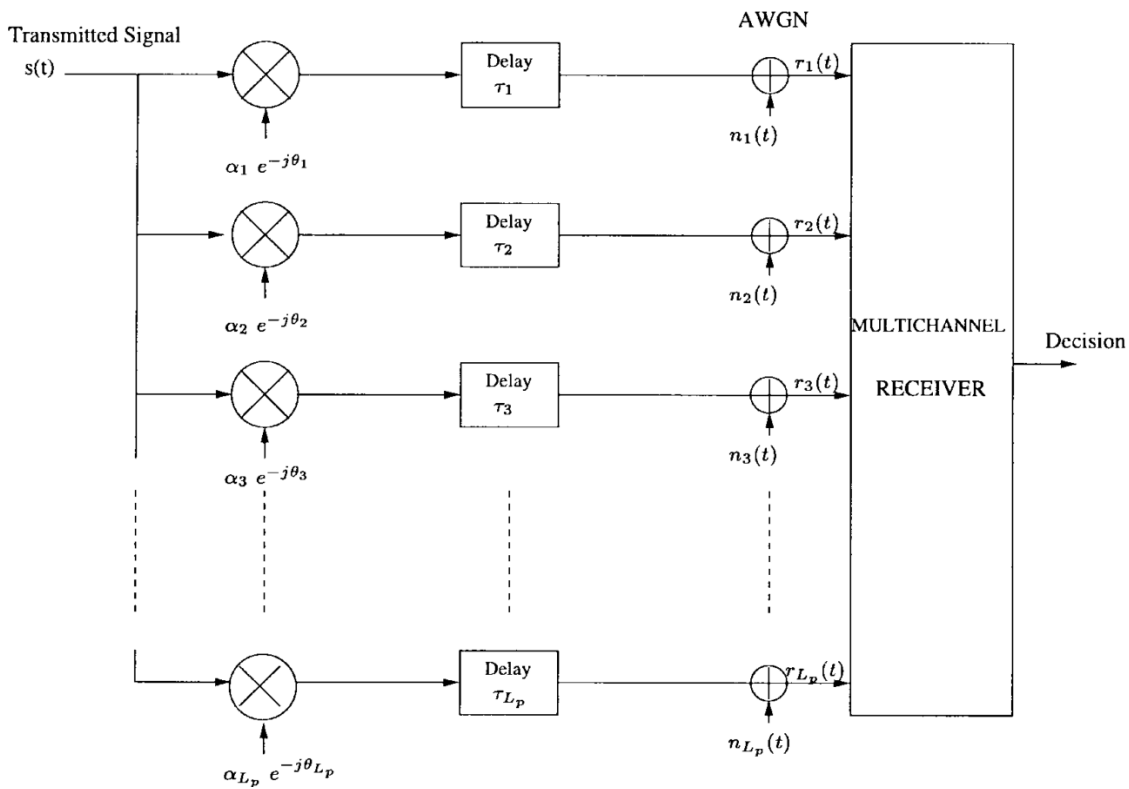

Figure 8 (Figure 1 of [42]): Diversity receiver analyzed for general modulations and channels.

The initial papers by Marv and Alouini on this topic relied on the MGF-approach to derive error probability expressions. Marv then started looking at other important performance metrics such as the outage probability. In particular, relying on a simple and accurate algorithm for the numerical inversion of the Laplace transforms of CDFs, Marv along with Alouini and Ko developed in [46] an MGF moment-based numerical technique for the outage probability evaluation of maximal-ratio combining and postdetection equal-gain combining in generalized fading channels for which the fading in each diversity path need not be independent, identically distributed, nor even distributed according to the same family of distributions. The method was then extended to coherent equal-gain combining (EGC) but only for the case of Nakagami-m fading channels. The large body of Marv's work with Alouini, culminating in their high-cited and comprehensive textbook [54], had significant impact on the field of error probability analysis.

Marv turned then his attention to the design and performance of coded modulation systems over fading channels. For instance, [55] derived the exact pairwise error probability for space-time coding over quasi-static Rayleigh fading channels in the presence of spatial fading correlation. In addition, the impact of correlation was quantified using the notion of "majorization". A follow-up paper [56] delved more into this topic and applied the MGF-based approach to derive analytical expressions for the exact pairwise error probability of a space-time coded system operating over spatially correlated fast (constant over the duration of a symbol) and slow (constant over the length of a code word) practical fading accounting for antenna spacing, antenna geometries and scattering models. Finally, in [57] a modification and a performance study of the traditional Alamouti code originally proposed for radio-frequency wireless applications was proposed in order to be applicable for on-off keying or pulse-position modulations which are popular in free-space optical communication with direct detection.

## 3.7 Software-Defined Radio

In the 1980s, Marv together with JPL colleagues Alexander Mileant and Sam Dolinar were working to develop SNR estimators and an analytical means to evaluate their performance [58-59]. In 2004, Marv and Dolinar began a collaboration to extend their previous SNR estimator results as part of a larger project to develop an autonomous SDR [60-62]. The objective of the autonomous radio project was to develop a radio receiver that could automatically configure itself via software to receive an incoming signal without much a priori knowledge of the defining characteristics of the signal. These characteristics include the carrier frequency and phase, modulation type, modulation index (ratio of transmitted carrier to data power, i.e. suppressed or non-suppressed carrier signal), symbol timing, data rate, type of channel coding, and signal-to-noise ratio. Traditional radio receivers are configured at the time of design for one or more types of modulation and coding, including different data rates. For example, the 802.11 WiFi standard has multiple modulation and coding types, and the type selected for a given transmission is conveyed to the radio receiver prior to data transmission. Thus, the receiver is informed of the modulation type, data rate, symbol time, and code rate prior to signal reception. This type of reconfiguration can be impractical and time-consuming, especially when the number of different modulations and codes becomes large, or when they have not even yet been designed. These difficulties motivate the need for an autonomous receiver that includes estimating and classifying modules to analyze the defining characteristics of the incoming signal and then to configure itself on the basis of the outputs of these modules.

The particular challenge addressed by Marv and Dolinar in [60-62] was to generalize existing SNR estimators such that they could operate when key aspects of the signal waveform are unknown. For example, most SNR estimators from the literature assumed that the modulation type, symbol rate, and pulse shape were known. The SNR estimator of [60-62] removed these assumptions. Specifically, it operated independently of the MPSK modulation order and carrier phase, providing a joint estimation of symbol rate, pulse shape, symbol synchronization, and SNR estimation. These SNR estimation results and related developments at JPL sparked much additional work by

Marv and others that culminated in Marv's last published book in 2006 [63], *Autonomous Software-Defined Radio Receivers for Deep Space Application*s, co-edited by Jon Hamkins and Marv. In addition to guiding the overall effort to get the book published, Marv and Hamkins ended up developing and analyzing most of the classification and estimation algorithms associated with an autonomous radio, which became key chapters in the book. In addition, the main ideas for autonomous radios were captured in U.S. patent 8,358,723 issued in 2013 [64]. The patent, entitled "Self-configurable radio receiver system and method for use with signals without prior knowledge of signal defining characteristics," had as co-inventors Marv and Hamkins, Divsalar, Dolinar, and Tkacenko, who were also the primary authors of the chapters in [63].

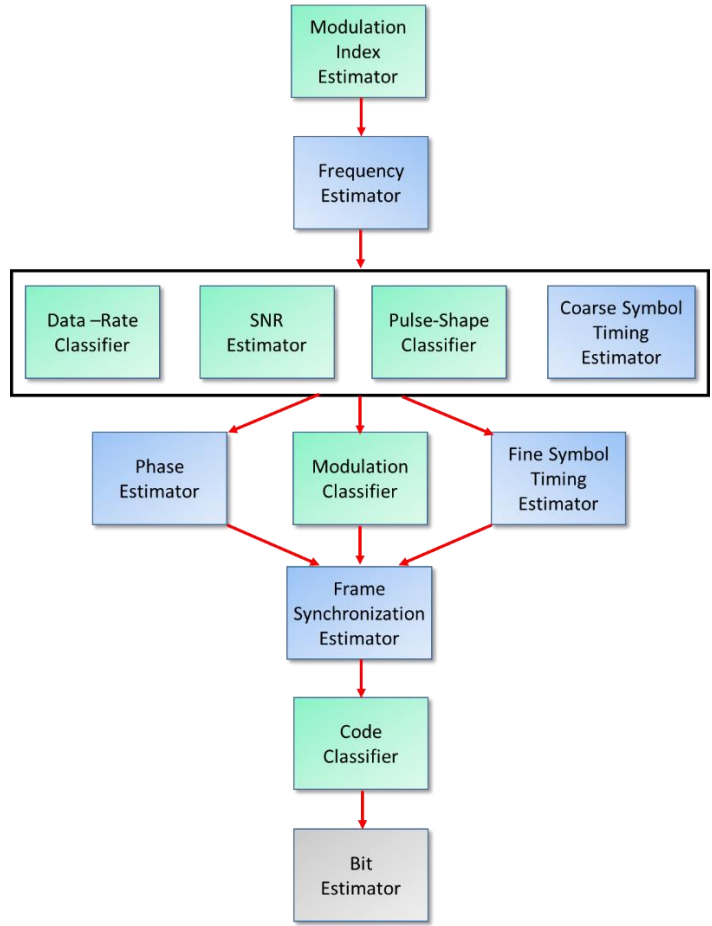

Figure 9: Estimation and classification modules of an autonomous radio. Order of modules and their dependencies are indicated in the diagram for the first iteration. After the first iteration, module output may be provided to other estimation modules to improve performance.

The proposed arrangement of classification and estimation modules of an autonomous radio is given in the partially ordered set shown in Figure 9. The modules in the figure include the frequency, phase, symbol timing, frame synchronization and bit estimation modules typical in traditional receiver architectures, shown in blue, as well as modules unique to the autonomous radio for estimation of modulation index, data rate, SNR, pulse shape, modulation type, and code parameters, shown in green. The bit estimation module in grey is a standard module applied once all signal parameters have been estimated. The solid arrows in the figure indicate dependencies where another module's input is required. These solid-line dependencies are critical; for example, it would not be feasible to classify the modulation type prior to classifying the data rate and obtaining the symbol timing. The cluster of classifiers/estimators for data rate, SNR, pulse shape, and coarse symbol timing are highly interdependent and the optimal technique provides joint estimation, although these modulates can operate independently [61, Chapter 7]. After the first estimate of the parameters in each module is obtained, the estimates may be fed laterally and upward to other estimation modules to improve performance. For example, the modulation classifier operates non-coherently at first, without input from the phase-tracking loop, but once a phase estimate is available, it may switch to a better-performing coherent modulation classifier. Similarly, the SNR estimator is designed to work optimally with or without side information about data rate, pulse shape, or symbol timing. This approach yields a bootstrapping approach to estimating/classifying all of the parameters necessary for the proper operation of the entire receiver.

At first glance, it may seem that the blue estimator modules of Figure 9 are simply long-established, conventional designs. For example, phase estimation via tracking loops have been designed and analyzed for nearly every reasonable signal type. However, the design of a phase-tracking loop for suppressed carrier signals typically assumes a known modulation type, yet for the autonomous radio, a loop is needed that works adequately for any phase-modulated signal, and which can improve its performance by later taking input from the modulation classifier when it starts producing an output. Similarly, the other seemingly standard modules have similar design challenges because of unknown signal attributes. Moreover, conventional implementations of frequency estimators, symbol-timing loops, and SNR estimators also assume the modulation type is known. In addition, there are a number of estimators that are not conventional and occur only in an autonomous radio, as shown in green in Figure 9. These include the blocks that estimate or classify the data rate, modulation index, and modulation type. Of these completely new modules, Marv developed or co-developed the algorithms and associated book chapters in [61] for the: a) modulation index estimation; b) data format and pulse shape classification; c) signal-to-noise ratio estimation; d) data rate estimation; e) carrier synchronization; f) modulation classification; and g) symbol synchronization. Some of these results also appeared in the literature prior to the book's publication [63-65]. In most cases these results involved formulating the maximum likelihood criterion for the estimator under the given unknowns and solving it analytically. This led to excellent solutions for modulation classification, SNR estimation, and frequency tracking. In some other cases, the maximum-likelihood solution was not tractable, and hence promising ad hoc schemes were identified, typically based on a mathematical analysis involving integration, non-linear operations, and appropriate approximations.

## 4  Closing Remarks

Communications technology continues to evolve at a rapid pace, building on many of the foundational ideas developed by Marv during his four decades of research. Indeed, many of the techniques Marv developed are experiencing a resurgence to address new challenges of emerging systems. In particular, millimeter wave communications requires far more demanding synchronization techniques than at lower frequencies, renewing the challenge of developing fast and accurate synchronization algorithms with low complexity. Diversity and range benefits of "Massive multiple-input multiple-output (MIMO) systems," whereby large antenna arrays are deployed at cellular base stations, are being studied and implemented for millimeter wave systems as well as at lower frequencies; Marv's diversity analysis techniques provide a powerful mechanism to understand the performance gains of such systems. The "Internet of Things" promises billions of additional devices sharing existing wireless spectrum, demanding new forms of multiple access and interference cancellation that will build on the groundbreaking work of Marv and others in spread spectrum. Finally, the standard for the next generation of cellular communication, 5G, is looking at new waveforms for data transmission, which will require the rigorous performance analysis techniques Marv developed for today's waveforms that helped cement them into existing systems. While Marv cannot help us solve these emerging challenges, his legacy of deep, rigorous, and creative research as outlined in this tribute will serve as an inspiration and model for the next generation of communication theorists in addressing emerging wireless system design challenges.

## Acknowledgements

The authors gratefully acknowledge Norm Beaulieu, Jon Hamkins, William Lindsey, Andreas Polydoros, and Robert Scholtz for their contributions to and comments on the tribute. They would also like to thank Marv's daughter Brette Simon for her comments as well as for sharing many of the pictures that appear in the tribute.